# Comment: Boosting Algorithms: Regularization, Prediction and Model Fitting

**Trevor Hastie**

We congratulate the authors (hereafter BH) for an interesting take on the boosting technology, and for developing a modular computational environment in R for exploring their models. Their use of low-degree-of-freedom smoothing splines as a base learner provides an interesting approach to adaptive additive modeling. The notion of "Twin Boosting" is interesting as well; besides the adaptive lasso, we have seen the idea applied more directly for the lasso and Dantzig selector (James, Radchenko and Lv (2007)).

In this discussion we elaborate on the connections between $L_2$-boosting of a linear model and infinitesimal forward stagewise linear regression. We then take the authors to task on their definition of degrees of freedom.

## 1. $L_2$-BOOST AND INFINITESIMAL FORWARD STAGEWISE LINEAR REGRESSION

Motivated by a version of $L_2$-boosting in Chapter 10 of Hastie, Tibshirani and Friedman (2001), Efron, Hastie, Johnstone and Tibshirani (2004) proposed the LARS algorithm. The intent was to:

- develop a limiting version of $L_2$-boost in which the step-length $\nu$ went to zero;
- show that this limiting version gave paths identical to the lasso, as was hinted in that chapter.

The result was three very similar varieties of the LARS algorithm, namely lasso, LAR and infinitesimal forward stagewise (iFSLR) (package lars for R, available from CRAN). iFSLR is indeed the limit of $L_2$-boost as $\nu \downarrow 0$, with piecewise-linear coefficient profiles, but is not always the same as the lasso.

*Trevor Hastie is Professor, Department of Statistics, Stanford University, Starford, California 94305, USA e-mail: hastie@stanford.edu.*

On a slight technical note, the version of $L_2$-boost proposed in BH is slightly different from that in Hastie, Tibshirani and Friedman (2001). Compare

$$(1) \quad \text{[BH]} \quad \hat{\beta}^{[m]} = \hat{\beta}^{[m-1]} + \nu \cdot \hat{\beta}^{(\hat{S}_m)},$$

$$(2) \quad \text{[HTF]} \quad \hat{\beta}^{[m]} = \hat{\beta}^{[m-1]} + \nu \cdot \text{sign}[\hat{\beta}^{(\hat{S}_m)}].$$

Despite the difference, they both have the same limit, which is computed exactly for squared-error loss by the type="forward.stagewise" option in the package lars. As $\nu$ gets very small, initially the same coefficient tends to get continuously updated by infinitesimal amounts (hence linearly). Eventually a second variable ties with the first for coefficient updates, which they share in a balanced way while remaining tied. Then a third joins in, and so on. Using simple least-squares computations, the LARS algorithm computes the entire iFSLR path with the same cost as a single multiple-least-squares fit. Note that in this limiting case, we can no longer index the sequence by step-number $m$ as in (1) or (2), but must resort to some other measure, such as the $L_1$-arc-length of the coefficient profile (Hastie, Taylor, Tibshirani and Walther (2007)).

Lasso and iFSLR are not always the same. In high-dimensional problems with correlated predictors, lasso profiles become wiggly quickly, whereas iFSLR profiles tend to be much smoother and monotone (Hastie et al. (2007)). Efron et al. (2004) establish sufficient *positive cone conditions* on the model matrix $X$ which effectively limit the amount of correlation between the variables and guarantee that lasso and iFSLR are the same; in particular, if the lasso profiles are monotone, all three algorithms are identical.

## 2. DEGREES OF FREEDOM

The authors propose a simple formula for the degrees of freedom for an $L_2$-boosted model. They construct the hat matrix $\mathcal{B}_m$ that computes the fit at iteration $m$, and then use $\mathrm{df}(m) = \mathrm{trace}(\mathcal{B}_m)$. They are in effect treating the model at stage $m$ as if it were computed by a predetermined sequence of linear updates. If this were the case, their formula would be spot on, by the accepted definitions for effective degrees of freedom for linear operators (Hastie et al., 2001; Efron et al., 2004). They acknowledge that this is an approximation (since the sequence was not predetermined, but rather adaptively chosen), but do not elaborate. In fact this approximation can be very badly off. Figure 1 shows the true degrees of freedom $\mathrm{df}_T(k)$ plotted against $\mathrm{df}(k)$ for two examples. We see that $\mathrm{df}(k)$ always underestimates $\mathrm{df}_T(k)$. We now discuss the details of these examples, and the basis for these claims.

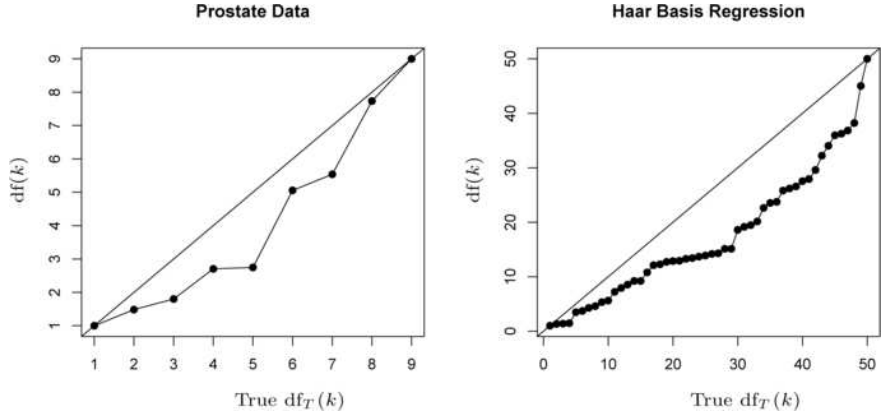

FIG. 1.  *The effective degrees of freedom for $L_2$-boost computed using the trace formula (vertical axis) vs. the exact degrees of freedom. The left plot is for the prostate cancer data example; the right plot is for a simulated univariate smoothing problem. In both cases df(m) underestimates the true degrees of freedom quite dramatically.*

The left example is the prostate data (Hastie et al., 2001, Figure 10.12) and has 67 observations and 9 predictors (including intercept). The right example fits a univariate piecewise-constant spline model of the form $f(x) = \sum_{j=1}^{50} \beta_j h_j(x)$, where the $h_j(x) = I(x \geq c_j)$ are a sequence of Haar basis functions with predefined knots $c_j$ at the unique values of the input values $x_i$. There are 50 observations and 50 predictors. In both problems we fit the limiting $L_2$-boost model iFSLR, using the `lars/forward.stagewise` procedure. Figure 2 shows the coefficient profiles.

In this case, using the results in Efron et al. (2004), it can be deduced that the equivalent limiting version of the hat matrix (5.6) of BH simplifies to a similar but more compact expression:

$$
\begin{aligned}
(3) \qquad \mathcal{B}_k = I &- (I - \gamma_k \mathcal{H}_k) \\
&\cdot (I - \gamma_{k-1}\mathcal{H}_{k-1}) \cdots (I - \gamma_1 \mathcal{H}_1).
\end{aligned}
$$

Here $k$ indexes the *step number* in the `lars` algorithm, where the steps delineate the breakpoints in the piecewise-linear path. $\mathcal{H}_j$ is the hat matrix corresponding to the variables involved in the $j$th portion of the piecewise linear path, and $\gamma_j$ is the relative distance in arc-length traveled along this piece until the next variable joins the active set (relative to the arc-length of the step that went all the way to the least squares fit). Using the BH definition, we would compute $\mathrm{df}(k) = \mathrm{trace}(\mathcal{B}_k)$ (vertical axis in Figure 1). 

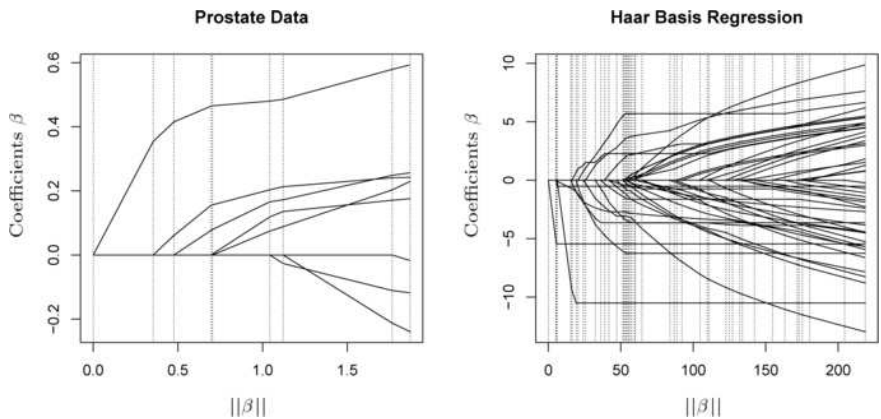

FIG. 2.  *Coefficient profiles for the iFSLR algorithm for the two examples. Both profiles are monotone, and are identical to the lasso profiles on these examples. In this case the df increment by 1 exactly at every vertical break-point line.*

These two examples were chosen carefully, for they both satisfy the *positive cone condition* mentioned above. In particular, the iFSLR path is the lasso path in both cases, and the active set grows by one at each step. More importantly, it is under these conditions that Efron et al. (2004) established that $\mathrm{df}_T(k) = k + 1$ *exactly* (horizontal axis in Figure 1). The $+1$ takes care of the intercept.

Consider the first step. The dominant variable enters the model, and gets its coefficient incremented until we reach the point that the next competitor is about to enter. At this point the df is exactly 2, while the formula $\mathrm{df}(1) = \mathrm{trace}(\mathcal{B}_1) = 1.48$ for the first example in Figure 1; this is off by 25%.

The exact df satisfies our intuition as well. If the first variable is far more significant than the rest, we will almost fit it entirely ($\gamma_1 \approx 1$) before the next one enters, and at that point the model has 2df. There is virtually no price for searching, because searching was not really needed. On the other hand, if many variables are competing for the first slot, shortly after the chosen one enters, another might appear, long before the first is fit completely ($\gamma_1 \ll 1$). Here the model also has 2df, despite the fact that the first variable has hardly progressed at all. This is the price paid for selection.

Even when the positive cone conditions are not satisfied, it can be shown that the size of the active set is an unbiased estimate of the true df (Zou, Hastie and Tibshirani (2007)).

It is possible that the authors can devise a correction for their df($k$) formula, based on the insights learned here. In some cases it may be possible to calibrate the formula to match the size of the active set. Failing that, one can use bootstrap methods to estimate df. But if the main purpose for estimating df is for model selection, K-fold cross-validation is a useful alternative.

## ACKNOWLEDGMENTS

This research was supported by NSF Grant DMS-05-05676 and NIH Grant 2R01 CA 72028-07.

## REFERENCES

EFRON, B., HASTIE, T., JOHNSTONE, I. and TIBSHIRANI, R. (2004). Least angle regression (with discussion). *Ann. Statist.* **32** 407–499. MR2060166

HASTIE, T., TAYLOR, J., TIBSHIRANI, R. and WALTHER, G. (2007). Forward stagewise regression and the monotone lasso. *Electron. J. Statist.* **1** 1–29. MR2312144

HASTIE, T., TIBSHIRANI, R. and FRIEDMAN, J. (2001). *The Elements of Statistical Learning. Data Mining, Inference, and Prediction.* Springer, New York. MR1851606

JAMES, G., RADCHENKO, P. and LV, J. (2007). The dasso algorithm for fitting the dantzig selector and the lasso. Technical report, Marshall School of Business, Univ. Southern California.

ZOU, H., HASTIE, T. and TIBSHIRANI, R. (2007). On the "degrees of freedom" of the lasso. *Ann. Statist.* **35** 2173–2192.